\documentclass[reprint, amsmath,amssymb, aps, superscriptaddress, floatfix]{revtex4-1}

\usepackage[utf8]{inputenc}
\usepackage{graphicx}
\usepackage{dcolumn}
\usepackage{bm}
\usepackage{braket}
\usepackage{color}
\usepackage{units}
\usepackage{dsfont}
\usepackage{float}
\usepackage{hyperref}
\usepackage{xcolor}
\usepackage{comment}

\begin{document}

\title{Terahertz electro-optic modulation of single photons}

\author{Nicolas~Couture}
\affiliation{National Research Council Canada, 100 Sussex Drive, Ottawa, Ontario K1N 5A2, Canada}
\affiliation{Department of Physics, University of Ottawa, Ottawa, Ontario K1N 6N5, Canada}

\author{Fr\'ed\'eric~Bouchard}
\email{frederic.bouchard@nrc-cnrc.gc.ca}
\affiliation{National Research Council Canada, 100 Sussex Drive, Ottawa, Ontario K1N 5A2, Canada}

\author{Alicia~Sit}
\affiliation{National Research Council Canada, 100 Sussex Drive, Ottawa, Ontario K1N 5A2, Canada}

\author{Guillaume~Thekkadath}
\affiliation{National Research Council Canada, 100 Sussex Drive, Ottawa, Ontario K1N 5A2, Canada}

\author{Duncan~England}
\affiliation{National Research Council Canada, 100 Sussex Drive, Ottawa, Ontario K1N 5A2, Canada}

\author{Philip~J.~Bustard}
\affiliation{National Research Council Canada, 100 Sussex Drive, Ottawa, Ontario K1N 5A2, Canada}

\author{Benjamin~J.~Sussman}
\affiliation{National Research Council Canada, 100 Sussex Drive, Ottawa, Ontario K1N 5A2, Canada}
\affiliation{Department of Physics, University of Ottawa, Ottawa, Ontario K1N 6N5, Canada}

\begin{abstract}
The manipulation of visible and near-infrared light at the single-photon level plays a key role in quantum communication systems where information is encoded into photonic degrees of freedom. In practical implementations, it is important to achieve this manipulation with high speeds, low loss, and low noise. In this work, we propose the use of terahertz~(THz) electric fields as a pump source for electro-optic modulation of single photons in bulk media. We demonstrate spectral modulations of single photons in the form of frequency translation and bandwidth manipulations as the terahertz field imparts linear and quadratic phases on the photons at various time delays within 1~ps. Our results show frequency translations exceeding 500~GHz at multi-THz modulation speeds and loss levels of $\approx$1~dB, complementing the current state-of-the-art electro-optic modulation techniques limited to speeds up to 100~GHz. The proposed approach leverages recent developments in terahertz generation techniques, introducing new avenues to manipulate non-classical light in an unexplored regime for quantum photonics.
\end{abstract}

\maketitle

The ability to manipulate light at the single-photon level on ultrafast timescales plays a key role in quantum photonic technologies~\cite{flamini_photonic_2018}, where high modulation speeds, large phase shifts, low losses, and low noise are required for functional applications. Ultrafast manipulation of single photons by three-wave mixing in materials with $\chi^{(2)}$ nonlinearity has led to a number of important demonstrations. A photon can be up-shifted by sum-frequency generation (SFG) with a strong pump~\cite{donohueUltrafast2014a}, down-shifted by difference-frequency generation (DFG), or compressed by chirped up-conversion~\cite{lavoieSpectral2013}. Quantum frequency conversion has also been shown to act as a mode-selective gate~\cite{ecksteinQuantum2011}. Equivalently, many of the same tasks can be performed by four-wave mixing in materials with $\chi^{(3)}$ nonlinearity including frequency translation by Bragg-scattering four-wave mixing~\cite{bonsma-fisherUltratunable2022a, mcguinnessQuantum2010a}, cross-phase modulation (XPM)~\cite{matsuda2016deterministic}, Raman quantum memory~\cite{fisher2016frequency}, molecular phase modulation~\cite{Tyumenev2022}, and all-optical gating via the optical Kerr effect~\cite{hall2011ultrafast,england_perspectives_2021}.

An attractive alternative to these nonlinear optical processing techniques is to use electronically driven lithium niobate-based electro-optic modulators (EOMs)~\cite{wright_spectral_2017,hiemstra_pure_2020,hu_-chip_2021,zhu_spectral_2022}. Importantly, the fast modulation of a phase imparted on an optical mode can result in the coherent manipulation of its spectral properties. In particular, linear temporal phase ramps lead to spectral translations, also known as spectral shears, whereas quadratic temporal phase ramps lead to spectral bandwidth compression or expansion. By altering the phase offset of a sinusoidal EOM field---thereby altering the time-varying phase profile applied to the single-photon-level pulses---these devices have achieved frequency translations of a few hundred~GHz and compression ratios above 400~\cite{zhu_spectral_2022,karpinski_bandwidth_2017,sosnicki_interface_2023,mittal_temporal_2017}.
However, compared to nonlinear optical modulation which can operate at $\gtrsim$THz modulation speeds, EOMs are currently limited to~$\sim$100~GHz, and in some cases loss in integrated devices can be prohibitively large. In this work we bridge the gap between nonlinear optical and electro-optical modulation by proposing, and demonstrating, the use of THz radiation for high-speed electro-optical modulation of quantum-level light.  

\begin{figure*}[ht!]
    \centering
    \includegraphics[width=2\columnwidth]{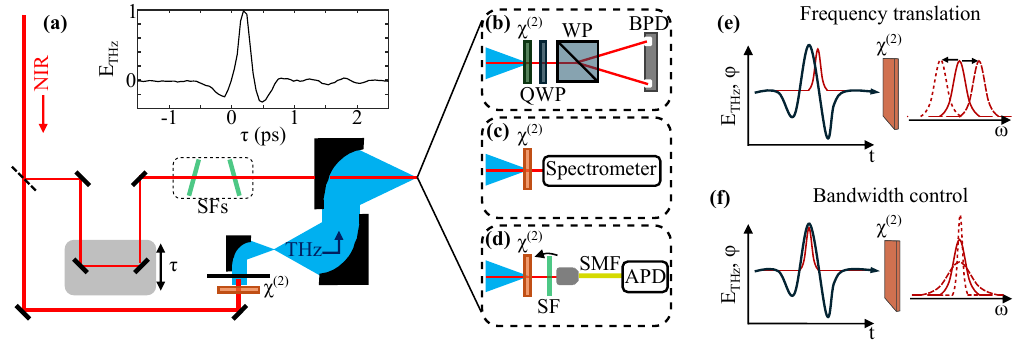}
    \caption[]{ (a) Schematic of the experimental setup and (inset) resulting terahertz electric field. The output from an ultrafast near-infrared source (red line) is split into two paths: one for the generation of pulsed terahertz radiation (blue line) and another for (b) terahertz detection through electro-optic sampling in a ZnTe crystal, (c) terahertz spectral modulation of classical light in a BNA crystal, or (d) terahertz spectral modulation of single photons in a BNA crystal. The oscillating terahertz electric field~($E_{\text{THz}}$) applies a time-varying phase~($\varphi$) on the near-infrared pulse through the electro-optic effect, a $\chi^{(2)}$ nonlinear process, where (e) linear and (f) quadratic phase profiles result in frequency translation and bandwidth control, respectively. SF: spectral filter, QWP: quarter-wave plate, WP: Wollaston prism, BPD: balanced photodiodes, SMF: single-mode fiber, APD: avalanche photodiode.}
    \label{fig:concept}
\end{figure*}

Achieving electro-optic phase modulation of single photons in bulk media has the potential for high-speed, low-loss modulation. Organic crystals have recently been identified as promising tools for the terahertz~(THz) community, due to their efficient emission of terahertz radiation~\cite{mansourzadeh_towards_2023}. Owing to their high $\chi^{(2)}$ coefficients and favorable phase-matching conditions---where the group velocity in the near-infrared regime approaches the phase velocity in the terahertz regime---optically pumping these crystals can produce extremely intense and broadband terahertz-field transients, spanning multiple octaves in a single cycle. The organic crystal N-benzyl-2-methyl-4-nitroaniline~(BNA) is regularly implemented as a high-field terahertz emitter~\cite{shalaby_demonstration_2015,shalaby_intense_2016,shalaby_extreme_2017,roeder_thz_2020}. However, the use of intense terahertz fields has yet to be used as a pump source for the electro-optic phase modulation of single photons, and has only recently become a frequency range of interest for quantum optical applications~\cite{kutas_terahertz_2020,kutas_quantum-inspired_2021,fandio_zeptojoule_2024,groiseau_single-photon_2024,kutas_terahertz_2024}. This approach holds great potential as terahertz fields routinely reach modulation speeds beyond the limit of electronics, while leveraging nonlinear effects in bulk media offers considerably lower insertion loss when compared to waveguide-based modulators.

In this work, we use terahertz electric fields and electro-optic crystals to manipulate single photons on ultrafast timescales. Organic BNA crystals are used first to produce high field-strength terahertz pulses, and subsequently act as a nonlinear medium for electro-optic phase modulation of single near-infrared~(NIR) photons centered at a carrier wavelength of 800\,nm~(375\,THz in frequency). By simply altering the delay between the terahertz field transient and single-photon-level near-infrared pulse, time-varying phase shifts are applied to the single photons to realize frequency translation, bandwidth compression, and bandwidth expansion. We demonstrate multi-THz phase modulation speeds and insertion loss of 1~dB, highlighting the viability of using terahertz pulses in future quantum networks that require ultrafast phase modulation and low loss~\cite{kimble_quantum_2008}.\\
\indent To demonstrate ultrafast electro-optic phase modulation of single photons using terahertz electric fields, these experiments rely on an amplified Ti:Sapphire laser delivering 30~fs pulses at a repetition rate of 100~Hz centered at a wavelength of 800~nm. The experimental configuration is shown in Fig.~\ref{fig:concept}a, and the various detection schemes used throughout this work are shown in Figs.~\ref{fig:concept}b-d. Terahertz generation is achieved via intra-pulse DFG in a 0.3~mm-thick BNA crystal. Based on the reported near-infrared-to-THz conversion efficiency between 0.2\% and 0.25\% in BNA~\cite{roeder_thz_2020,shalaby_intense_2016}, the highest peak terahertz electric field strength used in this work is estimated to be on the order of a few hundred kV/cm. The resulting terahertz electric field transient is characterized via electro-optic sampling in a zinc telluride (ZnTe) crystal~(Fig.~\ref{fig:concept}b) and plotted in Fig~\ref{fig:concept}a~(inset). A small portion of the near-infrared beam, referred to as the signal throughout the manuscript, is spectrally filtered to a bandwidth of $\sim$7~nm, appropriately attenuated, and passed through a variable delay line before being overlapped with the terahertz pulse inside a second nonlinear crystal. The terahertz-induced phase shift ($\Gamma$) on the signal photon is given by~\cite{planken2001measurement},
\begin{equation}
\label{eq:phase shift}
    \Gamma \propto \frac{\chi^{(2)}n_{\text{NIR}}^3E_{\text{THz}}\omega_{\text{NIR}}L}{c},
\end{equation}
where $n_{\text{NIR}}$ is the refractive index of the material at near-infrared frequencies, $E_{\text{THz}}$ is the terahertz electric field strength, $\omega_{\text{NIR}}$ is the near-infrared angular frequency, $c$ is the speed of light in vacuum, and $L$ is the crystal length. This can be maximized by replacing the ZnTe detection crystal used for electro-optic sampling with another BNA detection crystal, nearly identical to the one used for terahertz generation. Here, the terahertz pulse and the near-infrared photon are co-polarized and aligned parallel to the BNA crystal axis. In addition to again taking advantage of the high nonlinearity of BNA, $\chi^{(2)}_{\text{BNA}}\propto d_{\text{eff}}^{\text{BNA}}=234$~pm/V~\cite{fujiwara_determination_2007}, its dispersive properties result in a group velocity mismatch between the 800\,nm pump and the generated THz field which approaches zero at a frequency of 1.4~THz, thereby minimizing the temporal walk-off inside the crystal~(see Supplementary Material). By simply varying the delay between terahertz and near-infrared pulses interacting in detection crystal, we can control the time-varying phase shift that the near-infrared photons will experience, as depicted in Fig.~\ref{fig:concept}e~and~f. The delay can be set such that a linear or quadratic phase profile is applied to the single photons, resulting in frequency translation or bandwidth manipulation, respectively. This change in spectral properties is monitored as a measure of the terahertz-induced electro-optic phase modulations.
\begin{figure*}[ht!]
    \centering
    \includegraphics[width=2\columnwidth]{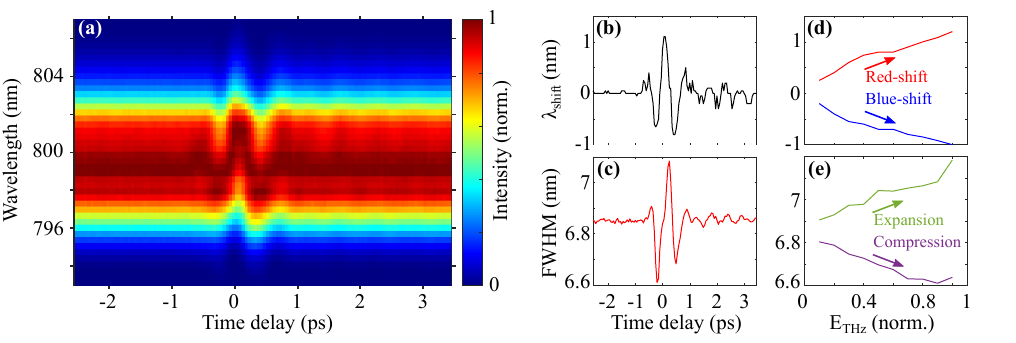}
    \caption[]{\textbf{Terahertz-induced spectral changes.} (a) Near-infrared spectra acquired as the relative delay between terahertz and near-infrared pulses is varied. (b) Relative shift in wavelength relative to the unmodulated spectrum and (c) spectral full-width-at-half maximum~(FWHM) bandwidth of the near-infrared pulse extracted from Gaussian fits applied to the raw spectra at each time delay shown in (a). (d) Maximum blue- and red-shifts and (e) spectral FWHM-bandwidth as the terahertz field strength ($E_{\text{THz}}$) is varied.}
    \label{fig:Classicalshear}
\end{figure*}\\
\indent We first characterize our system classically where the signal beam has high enough intensity such that the spectrum can be monitored with a conventional spectrometer, as depicted in Fig.~\ref{fig:concept}c. The signal spectrum is monitored as the relative delay~($\tau$) between signal and terahertz pulses is scanned; the normalized spectra are plotted in Fig.~\ref{fig:Classicalshear}a. The spectra are fitted to Gaussian profiles, and the shifted central wavelength and spectral bandwidth~(FWHM) are extracted and plotted in Fig.~\ref{fig:Classicalshear}b and c, respectively. The single-cycle terahertz electric field, oscillating over approximately 1~ps, achieves a maximum red-shift of approximately 1.2~nm~(560~GHz) and blue-shift of 1~nm~(470~GHz) relative to the central wavelength of the unmodulated signal beam at $\tau$=-2~ps. The frequency translation profile mostly follows the time derivative of the terahertz electric field transient shown in Fig.~\ref{fig:concept}a~(inset). The differing positive and negative terahertz electric field strengths results in differing absolute blue- and red-shifts in the near-infrared spectrum. Note that rotating the BNA crystal responsible for terahertz generation would flip the sign of the terahertz electric field and instead result in two red-shifts and one blue-shift. The shape of the terahertz field transient also adds a level of versatility to this approach, where multi-cycle pulses can be used to provide several blue- and red-shifts in the near-infrared spectrum over only a few picoseconds. Schemes utilizing XPM are often limited to a single blue- and red-shift due to the Gaussian temporal-intensity profile of ultrashort pulses~\cite{zhou_broadband_2020}. The shifts in the spectrum, shown in Fig.~\ref{fig:Classicalshear}d, increase linearly with $E_{\text{THz}}$, as is expected from Eq.~\ref{eq:phase shift}. Here, $E_{\text{THz}}=1$ corresponds to a near-infrared pulse energy of 100~$\mu$J dedicated to terahertz generation and few hundred kV/cm terahertz field strength.\\ 
\indent The spectral bandwidth reaches its highest modulation at $\tau$=100~fs~(Fig.~\ref{fig:Classicalshear}c), corresponding to a frequency translation per unit bandwidth of 17\%. This relative frequency shift is limited by the available terahertz field strengths, which are modest compared to the extreme terahertz fields capable of shifting near-infrared spectra by $>$100~nm~\cite{giorgianni_supercontinuum_2019}. In addition to frequency translation, increasing $E_{\text{THz}}$ offers higher magnitude bandwidth compression and expansion, as displayed by the results in Fig.~\ref{fig:Classicalshear}e. Although terahertz fields provide shorter time-lensing apertures than EOM-based approaches~\cite{zhu_spectral_2022,sosnicki_interface_2023}, appropriate tailoring of the signal spectrum and chirp combined with elevated terahertz fields can produce comparable compression and expansion ratios. The observed spectral modulation suggests that the $\sim$130-fs duration of the signal pulse is too long to experience purely linear or quadratic terahertz-induced phase profiles, causing the two processes to overlap. By omitting spectral filtering of the signal beam, these effects can be mitigated. However, the choice of a 130-fs pulse duration remains advantageous because it results in an appreciable shift when expressed as a percentage of the bandwidth.

\begin{figure*}[ht!]
    \centering
    \includegraphics[width=2\columnwidth]{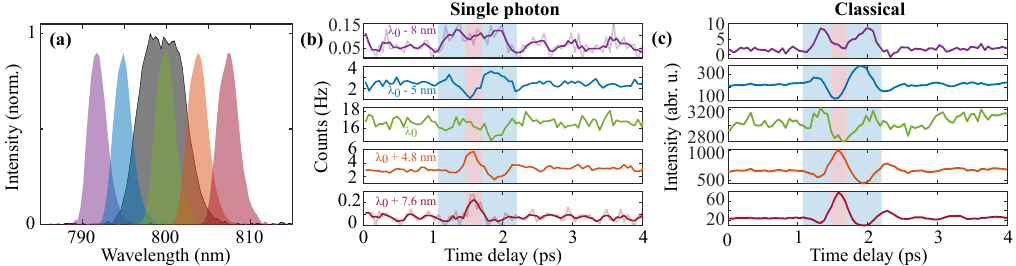}
     \caption[]{\textbf{Frequency translation of single photons.} (a)~Signal spectrum~(gray) and five spectral bands of $\sim$3-nm width explored for single photon experiments. (b)~Counts within 1-ns correlation window as a function of delay in each of the spectral bands presented in (a), where the shaded areas represent blue- and red-shifts in the single photon spectrum. A near-infrared pulse energy of 90~$\mu$J~(i.e. $E_{\text{THz}}=0.9$) was dedicated to terahertz generation for these measurements. (c) Spectra acquired in the classical regime (Fig.~\ref{fig:Classicalshear}) integrated over the spectral bands in (a).}
    \label{fig:shear}
\end{figure*}
\indent We also perform a version of the experiment at the single-photon level. The signal beam is strongly attenuated down to a mean photon number of 0.9 photons per pulse after considering the overall collection efficiency of the detection scheme, which encompasses the losses of all optical components and the efficiency of the detector. Notably, the loss contribution of the BNA crystal, where the electro-optic modulation occurs, is only 20\%~(or 1~dB) from Fresnel reflection at the front and back surfaces. Standard anti-reflection (AR) coating can be deposited on the back surface of the crystal to halve this loss without affecting the THz power inside the crystal, while more complex AR coating could further reduce the loss. To detect shifts of the single photon spectrum, a bandpass filter with 3-nm width is placed in front of the collection fiber, as depicted in Fig.~\ref{fig:concept}d, and angle-tuned to create five spectral bands~(Fig.~\ref{fig:shear}a). One band is centered at the signal central wavelength $\lambda_0=$~800~nm~(green), two are centered on the short-wavelength side of the spectrum at 792~nm and 795~nm~(violet and blue), and the remaining two are centered on the long-wavelength side of the spectrum at 804.8~nm and 807.6~nm~(orange and red). As the relative delay between signal and terahertz pulses is varied, in 50~fs steps, we observe the spectrum entering and exiting these spectral bands. The photons are measured with an avalanche photodiode~(APD) and time-tagging module, where a detection window of 1~ns about the expected arrival time of the photons is employed. A portion of the signal beam prior to attenuation is used as a trigger for the time-tagger. The count rate in each spectral band as a function of relative delay is plotted in Fig.~\ref{fig:shear}b. The spectral modulations measured in the single-photon regime are comparable to those observed with a significantly brighter signal in Fig~\ref{fig:Classicalshear}a and shown in Fig.~\ref{fig:shear}c, albeit with increased noise due to the uncertainties arising from the counting statistics. As expected the phase shift imparted onto the signal is independent of its strength~(Eq~\ref{eq:phase shift}), which permits the spectral modulation to be accurately characterized with classical-level pulses. Although we use weak coherent states as our single-photon source, electro-optic modulation is achievable with true single photon sources relying on spontaneous processes like SPDC~\cite{shields_electro-optical_2022}.\\
\indent Similar to the classical case shown in Fig.~\ref{fig:Classicalshear}d, the red-shift in the single photon spectrum is monitored as $E_{\text{THz}}$ is varied. Here, the spectral filter in front of the single-mode fiber is angle tuned to the orange band shown in Fig.~\ref{fig:shear}a and $\tau$ is held fixed to the delay corresponding to the highest count rate in Fig.~\ref{fig:shear}e. As $E_{\text{THz}}$ is increased, the count rate grows proportionally, indicating that the single photon spectrum is translating into the given spectral band through electro-optic modulation.\\

\begin{figure}[ht!]
    \centering
    \includegraphics[width=1\columnwidth]{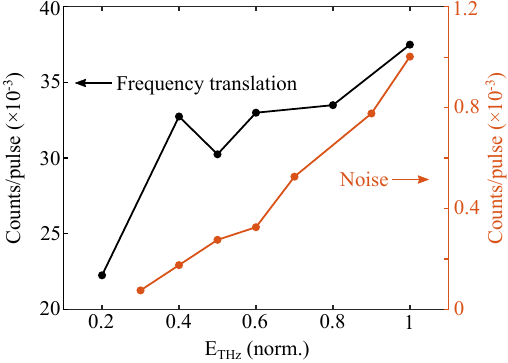}
    \caption{Counts per pulse~(black line, left axis) of the red-shifted spectrum in the orange band shown in Fig.~\ref{fig:shear}a~($\lambda_0+4.8$~nm) and noise counts per pulse~(orange line, right axis) as a function of terahertz electric field strength~($E_{\text{THz}}$).}
    \label{fig:ShearandNoise}
\end{figure}

\indent Performing the experiment at the single-photon level also allows us to carefully characterize the noise properties of our system, see Fig.~\ref{fig:shear}b. At the given terahertz field strength, corresponding to 90~$\mu$J of optical pump pulse energy dedicated to the terahertz generation process, the input signal field is blocked and noise counts on the order of 10$^{-4}$ per pulse are measured within the same correlation window as the signal. Noise counts are measured as a function of $E_{\text{THz}}$~(orange line in Fig.~\ref{fig:ShearandNoise}). Conventional single photon detectors are blind to THz frequencies, so that no THz filtering is required. Due to the low noise rate, and 100\,Hz repetition frequency we measure only $\sim5$ noise photons per second. Accordingly, the precise origin of the noise is hard to ascertain. We expect that the SNR would be significantly improved by selecting a signal wavelength that is spectrally removed from the noise associated with the strong 800\,nm pump pulses. 

\indent The approach we have presented in this work can benefit quantum photonic platforms requiring fast modulation of single photons with low loss and low noise. Contrary to techniques relying on sinusoidal electrical signals in EOMs, our approach leverages single-cycle terahartz fields with pulse duration on the order of 1~ps. Photons arriving before or after this window can pass through the system unaffected by the terahertz light, allowing information to be encoded into time bins separated by more than 1~ps. The large bandwidths of terahertz pulses also allow for arbitrary waveform shaping of the terahertz fields~\cite{Gingras:17,Gingras:18}, enabling a wide range of potential applications. Moreover, we have presented a low-noise alternative to optical XPM-based techniques that are burdened with parasitic noise from unwanted nonlinear effects from the strong optical pump pulses, without sacrificing modulation speed. Technological advancements promise to improve the practicality of our technique. For instance, high-field terahertz sources at multi-kHz repetition rates utilizing highly stable ytterbium-based ultrafast sources~\cite{cui_high-field_2023}, as opposed to the 100~Hz Ti:Sapphire source used here, would improve overall count rates of the system and reduce the noise level even further. The choice of nonlinear material for electro-optic modulation also adds a layer of practical control, which can be chosen, or even engineered, to satisfy phase-matching conditions and minimize group velocity mismatch at other infrared wavelengths of interest. Finally, this work marks an important step towards incorporating terahertz fields and sub-cycle pulses into the realm of quantum optics, inspiring new technologies or even enhancing existing architectures. \\

\begin{acknowledgments}
All authors thank Rune~Lausten, Denis~Guay, and Doug~Moffatt for technical support. All authors thank  David Purschke, Kate~Fenwick, Ramy~Tannous, Yingwen~Zhang, Andrew~Proppe, Noah~Lupu-Gladstein, Aaron~Goldberg, and Khabat~Heshami for useful discussions. All authors thank Swiss Terahertz LLC for providing BNA crystals.    
\end{acknowledgments}


\begin{thebibliography}{37}%
\makeatletter
\providecommand \@ifxundefined [1]{%
 \@ifx{#1\undefined}
}%
\providecommand \@ifnum [1]{%
 \ifnum #1\expandafter \@firstoftwo
 \else \expandafter \@secondoftwo
 \fi
}%
\providecommand \@ifx [1]{%
 \ifx #1\expandafter \@firstoftwo
 \else \expandafter \@secondoftwo
 \fi
}%
\providecommand \natexlab [1]{#1}%
\providecommand \enquote  [1]{``#1''}%
\providecommand \bibnamefont  [1]{#1}%
\providecommand \bibfnamefont [1]{#1}%
\providecommand \citenamefont [1]{#1}%
\providecommand \href@noop [0]{\@secondoftwo}%
\providecommand \href [0]{\begingroup \@sanitize@url \@href}%
\providecommand \@href[1]{\@@startlink{#1}\@@href}%
\providecommand \@@href[1]{\endgroup#1\@@endlink}%
\providecommand \@sanitize@url [0]{\catcode `\\12\catcode `\$12\catcode
  `\&12\catcode `\#12\catcode `\^12\catcode `\_12\catcode `\%12\relax}%
\providecommand \@@startlink[1]{}%
\providecommand \@@endlink[0]{}%
\providecommand \url  [0]{\begingroup\@sanitize@url \@url }%
\providecommand \@url [1]{\endgroup\@href {#1}{\urlprefix }}%
\providecommand \urlprefix  [0]{URL }%
\providecommand \Eprint [0]{\href }%
\providecommand \doibase [0]{http://dx.doi.org/}%
\providecommand \selectlanguage [0]{\@gobble}%
\providecommand \bibinfo  [0]{\@secondoftwo}%
\providecommand \bibfield  [0]{\@secondoftwo}%
\providecommand \translation [1]{[#1]}%
\providecommand \BibitemOpen [0]{}%
\providecommand \bibitemStop [0]{}%
\providecommand \bibitemNoStop [0]{.\EOS\space}%
\providecommand \EOS [0]{\spacefactor3000\relax}%
\providecommand \BibitemShut  [1]{\csname bibitem#1\endcsname}%
\let\auto@bib@innerbib\@empty
\bibitem [{\citenamefont {Flamini}\ \emph {et~al.}(2018)\citenamefont
  {Flamini}, \citenamefont {Spagnolo},\ and\ \citenamefont
  {Sciarrino}}]{flamini_photonic_2018}%
  \BibitemOpen
  \bibfield  {author} {\bibinfo {author} {\bibfnamefont {F.}~\bibnamefont
  {Flamini}}, \bibinfo {author} {\bibfnamefont {N.}~\bibnamefont {Spagnolo}}, \
  and\ \bibinfo {author} {\bibfnamefont {F.}~\bibnamefont {Sciarrino}},\ }\href
  {\doibase 10.1088/1361-6633/aad5b2} {\bibfield  {journal} {\bibinfo
  {journal} {Rep. Prog. Phys.}\ }\textbf {\bibinfo {volume} {82}},\ \bibinfo
  {pages} {016001} (\bibinfo {year} {2018})}\BibitemShut {NoStop}%
\bibitem [{\citenamefont {Donohue}\ \emph {et~al.}(2014)\citenamefont
  {Donohue}, \citenamefont {Lavoie},\ and\ \citenamefont
  {Resch}}]{donohueUltrafast2014a}%
  \BibitemOpen
  \bibfield  {author} {\bibinfo {author} {\bibfnamefont {J.~M.}\ \bibnamefont
  {Donohue}}, \bibinfo {author} {\bibfnamefont {J.}~\bibnamefont {Lavoie}}, \
  and\ \bibinfo {author} {\bibfnamefont {K.~J.}\ \bibnamefont {Resch}},\ }\href
  {\doibase 10.1103/PhysRevLett.113.163602} {\bibfield  {journal} {\bibinfo
  {journal} {Physical Review Letters}\ }\textbf {\bibinfo {volume} {113}},\
  \bibinfo {pages} {163602} (\bibinfo {year} {2014})}\BibitemShut {NoStop}%
\bibitem [{\citenamefont {Lavoie}\ \emph {et~al.}(2013)\citenamefont {Lavoie},
  \citenamefont {Donohue}, \citenamefont {Wright}, \citenamefont {Fedrizzi},\
  and\ \citenamefont {{K.J. Resch}}}]{lavoieSpectral2013}%
  \BibitemOpen
  \bibfield  {author} {\bibinfo {author} {\bibfnamefont {J.}~\bibnamefont
  {Lavoie}}, \bibinfo {author} {\bibfnamefont {J.~M.}\ \bibnamefont {Donohue}},
  \bibinfo {author} {\bibfnamefont {L.~G.}\ \bibnamefont {Wright}}, \bibinfo
  {author} {\bibfnamefont {A.}~\bibnamefont {Fedrizzi}}, \ and\ \bibinfo
  {author} {\bibnamefont {{K.J. Resch}}},\ }\href@noop {} {\bibfield  {journal}
  {\bibinfo  {journal} {Nature Photonics}\ }\textbf {\bibinfo {volume} {7}},\
  \bibinfo {pages} {363} (\bibinfo {year} {2013})}\BibitemShut {NoStop}%
\bibitem [{\citenamefont {Eckstein}\ \emph {et~al.}(2011)\citenamefont
  {Eckstein}, \citenamefont {Brecht},\ and\ \citenamefont
  {Silberhorn}}]{ecksteinQuantum2011}%
  \BibitemOpen
  \bibfield  {author} {\bibinfo {author} {\bibfnamefont {A.}~\bibnamefont
  {Eckstein}}, \bibinfo {author} {\bibfnamefont {B.}~\bibnamefont {Brecht}}, \
  and\ \bibinfo {author} {\bibfnamefont {C.}~\bibnamefont {Silberhorn}},\
  }\href {\doibase 10.1364/OE.19.013770} {\bibfield  {journal} {\bibinfo
  {journal} {Optics Express}\ }\textbf {\bibinfo {volume} {19}},\ \bibinfo
  {pages} {13770} (\bibinfo {year} {2011})}\BibitemShut {NoStop}%
\bibitem [{\citenamefont {{Bonsma-Fisher}}\ \emph {et~al.}(2022)\citenamefont
  {{Bonsma-Fisher}}, \citenamefont {Bustard}, \citenamefont {Parry},
  \citenamefont {Wright}, \citenamefont {England}, \citenamefont {Sussman},\
  and\ \citenamefont {Mosley}}]{bonsma-fisherUltratunable2022a}%
  \BibitemOpen
  \bibfield  {author} {\bibinfo {author} {\bibfnamefont {K.~A.~G.}\
  \bibnamefont {{Bonsma-Fisher}}}, \bibinfo {author} {\bibfnamefont {P.~J.}\
  \bibnamefont {Bustard}}, \bibinfo {author} {\bibfnamefont {C.}~\bibnamefont
  {Parry}}, \bibinfo {author} {\bibfnamefont {T.~A.}\ \bibnamefont {Wright}},
  \bibinfo {author} {\bibfnamefont {D.~G.}\ \bibnamefont {England}}, \bibinfo
  {author} {\bibfnamefont {B.~J.}\ \bibnamefont {Sussman}}, \ and\ \bibinfo
  {author} {\bibfnamefont {P.~J.}\ \bibnamefont {Mosley}},\ }\href {\doibase
  10.1103/PhysRevLett.129.203603} {\bibfield  {journal} {\bibinfo  {journal}
  {Physical Review Letters}\ }\textbf {\bibinfo {volume} {129}},\ \bibinfo
  {pages} {203603} (\bibinfo {year} {2022})}\BibitemShut {NoStop}%
\bibitem [{\citenamefont {McGuinness}\ \emph {et~al.}(2010)\citenamefont
  {McGuinness}, \citenamefont {Raymer}, \citenamefont {McKinstrie},\ and\
  \citenamefont {Radic}}]{mcguinnessQuantum2010a}%
  \BibitemOpen
  \bibfield  {author} {\bibinfo {author} {\bibfnamefont {H.~J.}\ \bibnamefont
  {McGuinness}}, \bibinfo {author} {\bibfnamefont {M.~G.}\ \bibnamefont
  {Raymer}}, \bibinfo {author} {\bibfnamefont {C.~J.}\ \bibnamefont
  {McKinstrie}}, \ and\ \bibinfo {author} {\bibfnamefont {S.}~\bibnamefont
  {Radic}},\ }\href {\doibase 10.1103/PhysRevLett.105.093604} {\bibfield
  {journal} {\bibinfo  {journal} {Physical Review Letters}\ }\textbf {\bibinfo
  {volume} {105}},\ \bibinfo {pages} {093604} (\bibinfo {year}
  {2010})}\BibitemShut {NoStop}%
\bibitem [{\citenamefont {Matsuda}(2016)}]{matsuda2016deterministic}%
  \BibitemOpen
  \bibfield  {author} {\bibinfo {author} {\bibfnamefont {N.}~\bibnamefont
  {Matsuda}},\ }\href@noop {} {\bibfield  {journal} {\bibinfo  {journal} {Sci.
  Adv.}\ }\textbf {\bibinfo {volume} {2}},\ \bibinfo {pages} {e1501223}
  (\bibinfo {year} {2016})}\BibitemShut {NoStop}%
\bibitem [{\citenamefont {Fisher}\ \emph {et~al.}(2016)\citenamefont {Fisher},
  \citenamefont {England}, \citenamefont {MacLean}, \citenamefont {Bustard},
  \citenamefont {Resch},\ and\ \citenamefont {Sussman}}]{fisher2016frequency}%
  \BibitemOpen
  \bibfield  {author} {\bibinfo {author} {\bibfnamefont {K.~A.}\ \bibnamefont
  {Fisher}}, \bibinfo {author} {\bibfnamefont {D.~G.}\ \bibnamefont {England}},
  \bibinfo {author} {\bibfnamefont {J.-P.~W.}\ \bibnamefont {MacLean}},
  \bibinfo {author} {\bibfnamefont {P.~J.}\ \bibnamefont {Bustard}}, \bibinfo
  {author} {\bibfnamefont {K.~J.}\ \bibnamefont {Resch}}, \ and\ \bibinfo
  {author} {\bibfnamefont {B.~J.}\ \bibnamefont {Sussman}},\ }\href@noop {}
  {\bibfield  {journal} {\bibinfo  {journal} {Nature communications}\ }\textbf
  {\bibinfo {volume} {7}},\ \bibinfo {pages} {11200} (\bibinfo {year}
  {2016})}\BibitemShut {NoStop}%
\bibitem [{\citenamefont {Tyumenev}\ \emph {et~al.}(2022)\citenamefont
  {Tyumenev}, \citenamefont {Hammer}, \citenamefont {Joly}, \citenamefont
  {Russell},\ and\ \citenamefont {Novoa}}]{Tyumenev2022}%
  \BibitemOpen
  \bibfield  {author} {\bibinfo {author} {\bibfnamefont {R.}~\bibnamefont
  {Tyumenev}}, \bibinfo {author} {\bibfnamefont {J.}~\bibnamefont {Hammer}},
  \bibinfo {author} {\bibfnamefont {N.~Y.}\ \bibnamefont {Joly}}, \bibinfo
  {author} {\bibfnamefont {P.~S.~J.}\ \bibnamefont {Russell}}, \ and\ \bibinfo
  {author} {\bibfnamefont {D.}~\bibnamefont {Novoa}},\ }\href {\doibase
  10.1126/science.abn1434} {\bibfield  {journal} {\bibinfo  {journal}
  {Science}\ }\textbf {\bibinfo {volume} {376}},\ \bibinfo {pages} {621}
  (\bibinfo {year} {2022})}\BibitemShut {NoStop}%
\bibitem [{\citenamefont {Hall}\ \emph {et~al.}(2011)\citenamefont {Hall},
  \citenamefont {Altepeter},\ and\ \citenamefont {Kumar}}]{hall2011ultrafast}%
  \BibitemOpen
  \bibfield  {author} {\bibinfo {author} {\bibfnamefont {M.~A.}\ \bibnamefont
  {Hall}}, \bibinfo {author} {\bibfnamefont {J.~B.}\ \bibnamefont {Altepeter}},
  \ and\ \bibinfo {author} {\bibfnamefont {P.}~\bibnamefont {Kumar}},\
  }\href@noop {} {\bibfield  {journal} {\bibinfo  {journal} {Phys. Rev. Lett.}\
  }\textbf {\bibinfo {volume} {106}},\ \bibinfo {pages} {053901} (\bibinfo
  {year} {2011})}\BibitemShut {NoStop}%
\bibitem [{\citenamefont {England}\ \emph {et~al.}(2021)\citenamefont
  {England}, \citenamefont {Bouchard}, \citenamefont {Fenwick}, \citenamefont
  {Bonsma-Fisher}, \citenamefont {Zhang}, \citenamefont {Bustard},\ and\
  \citenamefont {Sussman}}]{england_perspectives_2021}%
  \BibitemOpen
  \bibfield  {author} {\bibinfo {author} {\bibfnamefont {D.}~\bibnamefont
  {England}}, \bibinfo {author} {\bibfnamefont {F.}~\bibnamefont {Bouchard}},
  \bibinfo {author} {\bibfnamefont {K.}~\bibnamefont {Fenwick}}, \bibinfo
  {author} {\bibfnamefont {K.}~\bibnamefont {Bonsma-Fisher}}, \bibinfo {author}
  {\bibfnamefont {Y.}~\bibnamefont {Zhang}}, \bibinfo {author} {\bibfnamefont
  {P.~J.}\ \bibnamefont {Bustard}}, \ and\ \bibinfo {author} {\bibfnamefont
  {B.~J.}\ \bibnamefont {Sussman}},\ }\href {\doibase 10.1063/5.0065222}
  {\bibfield  {journal} {\bibinfo  {journal} {Appl. Phys. Lett.}\ }\textbf
  {\bibinfo {volume} {119}},\ \bibinfo {pages} {160501} (\bibinfo {year}
  {2021})}\BibitemShut {NoStop}%
\bibitem [{\citenamefont {Wright}\ \emph {et~al.}(2017)\citenamefont {Wright},
  \citenamefont {Karpiński}, \citenamefont {Söller},\ and\ \citenamefont
  {Smith}}]{wright_spectral_2017}%
  \BibitemOpen
  \bibfield  {author} {\bibinfo {author} {\bibfnamefont {L.~J.}\ \bibnamefont
  {Wright}}, \bibinfo {author} {\bibfnamefont {M.}~\bibnamefont {Karpiński}},
  \bibinfo {author} {\bibfnamefont {C.}~\bibnamefont {Söller}}, \ and\
  \bibinfo {author} {\bibfnamefont {B.~J.}\ \bibnamefont {Smith}},\ }\href
  {\doibase 10.1103/PhysRevLett.118.023601} {\bibfield  {journal} {\bibinfo
  {journal} {Phys. Rev. Lett.}\ }\textbf {\bibinfo {volume} {118}},\ \bibinfo
  {pages} {023601} (\bibinfo {year} {2017})}\BibitemShut {NoStop}%
\bibitem [{\citenamefont {Hiemstra}\ \emph {et~al.}(2020)\citenamefont
  {Hiemstra}, \citenamefont {Parker}, \citenamefont {Humphreys}, \citenamefont
  {Tiedau}, \citenamefont {Beck}, \citenamefont {Karpiński}, \citenamefont
  {Smith}, \citenamefont {Eckstein}, \citenamefont {Kolthammer},\ and\
  \citenamefont {Walmsley}}]{hiemstra_pure_2020}%
  \BibitemOpen
  \bibfield  {author} {\bibinfo {author} {\bibfnamefont {T.}~\bibnamefont
  {Hiemstra}}, \bibinfo {author} {\bibfnamefont {T.~F.}\ \bibnamefont
  {Parker}}, \bibinfo {author} {\bibfnamefont {P.}~\bibnamefont {Humphreys}},
  \bibinfo {author} {\bibfnamefont {J.}~\bibnamefont {Tiedau}}, \bibinfo
  {author} {\bibfnamefont {M.}~\bibnamefont {Beck}}, \bibinfo {author}
  {\bibfnamefont {M.}~\bibnamefont {Karpiński}}, \bibinfo {author}
  {\bibfnamefont {B.~J.}\ \bibnamefont {Smith}}, \bibinfo {author}
  {\bibfnamefont {A.}~\bibnamefont {Eckstein}}, \bibinfo {author}
  {\bibfnamefont {W.~S.}\ \bibnamefont {Kolthammer}}, \ and\ \bibinfo {author}
  {\bibfnamefont {I.~A.}\ \bibnamefont {Walmsley}},\ }\href {\doibase
  10.1103/PhysRevApplied.14.014052} {\bibfield  {journal} {\bibinfo  {journal}
  {Phys. Rev. Appl.}\ }\textbf {\bibinfo {volume} {14}},\ \bibinfo {pages}
  {014052} (\bibinfo {year} {2020})}\BibitemShut {NoStop}%
\bibitem [{\citenamefont {Hu}\ \emph {et~al.}(2021)\citenamefont {Hu},
  \citenamefont {Yu}, \citenamefont {Zhu}, \citenamefont {Sinclair},
  \citenamefont {Shams-Ansari}, \citenamefont {Shao}, \citenamefont
  {Holzgrafe}, \citenamefont {Puma}, \citenamefont {Zhang},\ and\ \citenamefont
  {Lončar}}]{hu_-chip_2021}%
  \BibitemOpen
  \bibfield  {author} {\bibinfo {author} {\bibfnamefont {Y.}~\bibnamefont
  {Hu}}, \bibinfo {author} {\bibfnamefont {M.}~\bibnamefont {Yu}}, \bibinfo
  {author} {\bibfnamefont {D.}~\bibnamefont {Zhu}}, \bibinfo {author}
  {\bibfnamefont {N.}~\bibnamefont {Sinclair}}, \bibinfo {author}
  {\bibfnamefont {A.}~\bibnamefont {Shams-Ansari}}, \bibinfo {author}
  {\bibfnamefont {L.}~\bibnamefont {Shao}}, \bibinfo {author} {\bibfnamefont
  {J.}~\bibnamefont {Holzgrafe}}, \bibinfo {author} {\bibfnamefont
  {E.}~\bibnamefont {Puma}}, \bibinfo {author} {\bibfnamefont {M.}~\bibnamefont
  {Zhang}}, \ and\ \bibinfo {author} {\bibfnamefont {M.}~\bibnamefont
  {Lončar}},\ }\href {\doibase 10.1038/s41586-021-03999-x} {\bibfield
  {journal} {\bibinfo  {journal} {Nature}\ }\textbf {\bibinfo {volume} {599}},\
  \bibinfo {pages} {587} (\bibinfo {year} {2021})}\BibitemShut {NoStop}%
\bibitem [{\citenamefont {Zhu}\ \emph {et~al.}(2022)\citenamefont {Zhu},
  \citenamefont {Chen}, \citenamefont {Yu}, \citenamefont {Shao}, \citenamefont
  {Hu}, \citenamefont {Xin}, \citenamefont {Yeh}, \citenamefont {Ghosh},
  \citenamefont {He}, \citenamefont {Reimer}, \citenamefont {Sinclair},
  \citenamefont {Wong}, \citenamefont {Zhang},\ and\ \citenamefont
  {Lončar}}]{zhu_spectral_2022}%
  \BibitemOpen
  \bibfield  {author} {\bibinfo {author} {\bibfnamefont {D.}~\bibnamefont
  {Zhu}}, \bibinfo {author} {\bibfnamefont {C.}~\bibnamefont {Chen}}, \bibinfo
  {author} {\bibfnamefont {M.}~\bibnamefont {Yu}}, \bibinfo {author}
  {\bibfnamefont {L.}~\bibnamefont {Shao}}, \bibinfo {author} {\bibfnamefont
  {Y.}~\bibnamefont {Hu}}, \bibinfo {author} {\bibfnamefont {C.~J.}\
  \bibnamefont {Xin}}, \bibinfo {author} {\bibfnamefont {M.}~\bibnamefont
  {Yeh}}, \bibinfo {author} {\bibfnamefont {S.}~\bibnamefont {Ghosh}}, \bibinfo
  {author} {\bibfnamefont {L.}~\bibnamefont {He}}, \bibinfo {author}
  {\bibfnamefont {C.}~\bibnamefont {Reimer}}, \bibinfo {author} {\bibfnamefont
  {N.}~\bibnamefont {Sinclair}}, \bibinfo {author} {\bibfnamefont {F.~N.~C.}\
  \bibnamefont {Wong}}, \bibinfo {author} {\bibfnamefont {M.}~\bibnamefont
  {Zhang}}, \ and\ \bibinfo {author} {\bibfnamefont {M.}~\bibnamefont
  {Lončar}},\ }\href {\doibase 10.1038/s41377-022-01029-7} {\bibfield
  {journal} {\bibinfo  {journal} {Light Sci. Appl.}\ }\textbf {\bibinfo
  {volume} {11}},\ \bibinfo {pages} {327} (\bibinfo {year} {2022})}\BibitemShut
  {NoStop}%
\bibitem [{\citenamefont {Karpiński}\ \emph {et~al.}(2017)\citenamefont
  {Karpiński}, \citenamefont {Jachura}, \citenamefont {Wright},\ and\
  \citenamefont {Smith}}]{karpinski_bandwidth_2017}%
  \BibitemOpen
  \bibfield  {author} {\bibinfo {author} {\bibfnamefont {M.}~\bibnamefont
  {Karpiński}}, \bibinfo {author} {\bibfnamefont {M.}~\bibnamefont {Jachura}},
  \bibinfo {author} {\bibfnamefont {L.~J.}\ \bibnamefont {Wright}}, \ and\
  \bibinfo {author} {\bibfnamefont {B.~J.}\ \bibnamefont {Smith}},\ }\href
  {\doibase 10.1038/nphoton.2016.228} {\bibfield  {journal} {\bibinfo
  {journal} {Nat. Photon.}\ }\textbf {\bibinfo {volume} {11}},\ \bibinfo
  {pages} {53} (\bibinfo {year} {2017})}\BibitemShut {NoStop}%
\bibitem [{\citenamefont {Sośnicki}\ \emph {et~al.}(2023)\citenamefont
  {Sośnicki}, \citenamefont {Mikołajczyk}, \citenamefont {Golestani},\ and\
  \citenamefont {Karpiński}}]{sosnicki_interface_2023}%
  \BibitemOpen
  \bibfield  {author} {\bibinfo {author} {\bibfnamefont {F.}~\bibnamefont
  {Sośnicki}}, \bibinfo {author} {\bibfnamefont {M.}~\bibnamefont
  {Mikołajczyk}}, \bibinfo {author} {\bibfnamefont {A.}~\bibnamefont
  {Golestani}}, \ and\ \bibinfo {author} {\bibfnamefont {M.}~\bibnamefont
  {Karpiński}},\ }\href {\doibase 10.1038/s41566-023-01214-z} {\bibfield
  {journal} {\bibinfo  {journal} {Nat. Photon.}\ }\textbf {\bibinfo {volume}
  {17}},\ \bibinfo {pages} {761} (\bibinfo {year} {2023})}\BibitemShut
  {NoStop}%
\bibitem [{\citenamefont {Mittal}\ \emph {et~al.}(2017)\citenamefont {Mittal},
  \citenamefont {Orre}, \citenamefont {Restelli}, \citenamefont {Salem},
  \citenamefont {Goldschmidt},\ and\ \citenamefont
  {Hafezi}}]{mittal_temporal_2017}%
  \BibitemOpen
  \bibfield  {author} {\bibinfo {author} {\bibfnamefont {S.}~\bibnamefont
  {Mittal}}, \bibinfo {author} {\bibfnamefont {V.~V.}\ \bibnamefont {Orre}},
  \bibinfo {author} {\bibfnamefont {A.}~\bibnamefont {Restelli}}, \bibinfo
  {author} {\bibfnamefont {R.}~\bibnamefont {Salem}}, \bibinfo {author}
  {\bibfnamefont {E.~A.}\ \bibnamefont {Goldschmidt}}, \ and\ \bibinfo {author}
  {\bibfnamefont {M.}~\bibnamefont {Hafezi}},\ }\href {\doibase
  10.1103/PhysRevA.96.043807} {\bibfield  {journal} {\bibinfo  {journal} {Phys.
  Rev. A}\ }\textbf {\bibinfo {volume} {96}},\ \bibinfo {pages} {043807}
  (\bibinfo {year} {2017})}\BibitemShut {NoStop}%
\bibitem [{\citenamefont {Mansourzadeh}\ \emph {et~al.}(2023)\citenamefont
  {Mansourzadeh}, \citenamefont {Vogel}, \citenamefont {Omar}, \citenamefont
  {Buchmann}, \citenamefont {Kelleher}, \citenamefont {Jepsen},\ and\
  \citenamefont {Saraceno}}]{mansourzadeh_towards_2023}%
  \BibitemOpen
  \bibfield  {author} {\bibinfo {author} {\bibfnamefont {S.}~\bibnamefont
  {Mansourzadeh}}, \bibinfo {author} {\bibfnamefont {T.}~\bibnamefont {Vogel}},
  \bibinfo {author} {\bibfnamefont {A.}~\bibnamefont {Omar}}, \bibinfo {author}
  {\bibfnamefont {T.~O.}\ \bibnamefont {Buchmann}}, \bibinfo {author}
  {\bibfnamefont {E.~J.~R.}\ \bibnamefont {Kelleher}}, \bibinfo {author}
  {\bibfnamefont {P.~U.}\ \bibnamefont {Jepsen}}, \ and\ \bibinfo {author}
  {\bibfnamefont {C.~J.}\ \bibnamefont {Saraceno}},\ }\href {\doibase
  10.1364/OME.502209} {\bibfield  {journal} {\bibinfo  {journal} {Opt. Mater.
  Express}\ }\textbf {\bibinfo {volume} {13}},\ \bibinfo {pages} {3287}
  (\bibinfo {year} {2023})}\BibitemShut {NoStop}%
\bibitem [{\citenamefont {Shalaby}\ and\ \citenamefont
  {Hauri}(2015)}]{shalaby_demonstration_2015}%
  \BibitemOpen
  \bibfield  {author} {\bibinfo {author} {\bibfnamefont {M.}~\bibnamefont
  {Shalaby}}\ and\ \bibinfo {author} {\bibfnamefont {C.~P.}\ \bibnamefont
  {Hauri}},\ }\href {\doibase 10.1038/ncomms6976} {\bibfield  {journal}
  {\bibinfo  {journal} {Nat. Commun.}\ }\textbf {\bibinfo {volume} {6}},\
  \bibinfo {pages} {5976} (\bibinfo {year} {2015})}\BibitemShut {NoStop}%
\bibitem [{\citenamefont {Shalaby}\ \emph {et~al.}(2016)\citenamefont
  {Shalaby}, \citenamefont {Vicario}, \citenamefont {Thirupugalmani},
  \citenamefont {Brahadeeswaran},\ and\ \citenamefont
  {Hauri}}]{shalaby_intense_2016}%
  \BibitemOpen
  \bibfield  {author} {\bibinfo {author} {\bibfnamefont {M.}~\bibnamefont
  {Shalaby}}, \bibinfo {author} {\bibfnamefont {C.}~\bibnamefont {Vicario}},
  \bibinfo {author} {\bibfnamefont {K.}~\bibnamefont {Thirupugalmani}},
  \bibinfo {author} {\bibfnamefont {S.}~\bibnamefont {Brahadeeswaran}}, \ and\
  \bibinfo {author} {\bibfnamefont {C.~P.}\ \bibnamefont {Hauri}},\ }\href
  {\doibase 10.1364/OL.41.001777} {\bibfield  {journal} {\bibinfo  {journal}
  {Opt. Lett.}\ }\textbf {\bibinfo {volume} {41}},\ \bibinfo {pages} {1777}
  (\bibinfo {year} {2016})}\BibitemShut {NoStop}%
\bibitem [{\citenamefont {Shalaby}\ \emph {et~al.}(2017)\citenamefont
  {Shalaby}, \citenamefont {Vicario},\ and\ \citenamefont
  {Hauri}}]{shalaby_extreme_2017}%
  \BibitemOpen
  \bibfield  {author} {\bibinfo {author} {\bibfnamefont {M.}~\bibnamefont
  {Shalaby}}, \bibinfo {author} {\bibfnamefont {C.}~\bibnamefont {Vicario}}, \
  and\ \bibinfo {author} {\bibfnamefont {C.~P.}\ \bibnamefont {Hauri}},\ }\href
  {\doibase 10.1063/1.4978051} {\bibfield  {journal} {\bibinfo  {journal} {APL
  Photonics}\ }\textbf {\bibinfo {volume} {2}},\ \bibinfo {pages} {036106}
  (\bibinfo {year} {2017})}\BibitemShut {NoStop}%
\bibitem [{\citenamefont {Roeder}\ \emph {et~al.}(2020)\citenamefont {Roeder},
  \citenamefont {Shalaby}, \citenamefont {Beleites}, \citenamefont
  {Ronneberger},\ and\ \citenamefont {Gopal}}]{roeder_thz_2020}%
  \BibitemOpen
  \bibfield  {author} {\bibinfo {author} {\bibfnamefont {F.}~\bibnamefont
  {Roeder}}, \bibinfo {author} {\bibfnamefont {M.}~\bibnamefont {Shalaby}},
  \bibinfo {author} {\bibfnamefont {B.}~\bibnamefont {Beleites}}, \bibinfo
  {author} {\bibfnamefont {F.}~\bibnamefont {Ronneberger}}, \ and\ \bibinfo
  {author} {\bibfnamefont {A.}~\bibnamefont {Gopal}},\ }\href {\doibase
  10.1364/OE.404690} {\bibfield  {journal} {\bibinfo  {journal} {Opt. Express}\
  }\textbf {\bibinfo {volume} {28}},\ \bibinfo {pages} {36274} (\bibinfo {year}
  {2020})}\BibitemShut {NoStop}%
\bibitem [{\citenamefont {Kutas}\ \emph {et~al.}(2020)\citenamefont {Kutas},
  \citenamefont {Haase}, \citenamefont {Bickert}, \citenamefont {Riexinger},
  \citenamefont {Molter},\ and\ \citenamefont {von
  Freymann}}]{kutas_terahertz_2020}%
  \BibitemOpen
  \bibfield  {author} {\bibinfo {author} {\bibfnamefont {M.}~\bibnamefont
  {Kutas}}, \bibinfo {author} {\bibfnamefont {B.}~\bibnamefont {Haase}},
  \bibinfo {author} {\bibfnamefont {P.}~\bibnamefont {Bickert}}, \bibinfo
  {author} {\bibfnamefont {F.}~\bibnamefont {Riexinger}}, \bibinfo {author}
  {\bibfnamefont {D.}~\bibnamefont {Molter}}, \ and\ \bibinfo {author}
  {\bibfnamefont {G.}~\bibnamefont {von Freymann}},\ }\href {\doibase
  10.1126/sciadv.aaz8065} {\bibfield  {journal} {\bibinfo  {journal} {Sci.
  Adv.}\ }\textbf {\bibinfo {volume} {6}},\ \bibinfo {pages} {eaaz8065}
  (\bibinfo {year} {2020})}\BibitemShut {NoStop}%
\bibitem [{\citenamefont {Kutas}\ \emph {et~al.}(2021)\citenamefont {Kutas},
  \citenamefont {Haase}, \citenamefont {Klier}, \citenamefont {Molter},\ and\
  \citenamefont {Freymann}}]{kutas_quantum-inspired_2021}%
  \BibitemOpen
  \bibfield  {author} {\bibinfo {author} {\bibfnamefont {M.}~\bibnamefont
  {Kutas}}, \bibinfo {author} {\bibfnamefont {B.}~\bibnamefont {Haase}},
  \bibinfo {author} {\bibfnamefont {J.}~\bibnamefont {Klier}}, \bibinfo
  {author} {\bibfnamefont {D.}~\bibnamefont {Molter}}, \ and\ \bibinfo {author}
  {\bibfnamefont {G.~v.}\ \bibnamefont {Freymann}},\ }\href {\doibase
  10.1364/OPTICA.415627} {\bibfield  {journal} {\bibinfo  {journal} {Optica}\
  }\textbf {\bibinfo {volume} {8}},\ \bibinfo {pages} {438} (\bibinfo {year}
  {2021})}\BibitemShut {NoStop}%
\bibitem [{\citenamefont {Fandio}\ \emph {et~al.}(2024)\citenamefont {Fandio},
  \citenamefont {Vishnuradhan}, \citenamefont {Yalavarthi}, \citenamefont
  {Cui}, \citenamefont {Couture}, \citenamefont {Gamouras},\ and\ \citenamefont
  {Ménard}}]{fandio_zeptojoule_2024}%
  \BibitemOpen
  \bibfield  {author} {\bibinfo {author} {\bibfnamefont {D.~J.~J.}\
  \bibnamefont {Fandio}}, \bibinfo {author} {\bibfnamefont {A.}~\bibnamefont
  {Vishnuradhan}}, \bibinfo {author} {\bibfnamefont {E.~K.}\ \bibnamefont
  {Yalavarthi}}, \bibinfo {author} {\bibfnamefont {W.}~\bibnamefont {Cui}},
  \bibinfo {author} {\bibfnamefont {N.}~\bibnamefont {Couture}}, \bibinfo
  {author} {\bibfnamefont {A.}~\bibnamefont {Gamouras}}, \ and\ \bibinfo
  {author} {\bibfnamefont {J.-M.}\ \bibnamefont {Ménard}},\ }\href {\doibase
  10.1364/OL.517916} {\bibfield  {journal} {\bibinfo  {journal} {Opt. Lett.}\
  }\textbf {\bibinfo {volume} {49}},\ \bibinfo {pages} {1556} (\bibinfo {year}
  {2024})}\BibitemShut {NoStop}%
\bibitem [{\citenamefont {Groiseau}\ \emph {et~al.}(2024)\citenamefont
  {Groiseau}, \citenamefont {Fernández-Domínguez}, \citenamefont
  {Martín-Cano},\ and\ \citenamefont {Muñoz}}]{groiseau_single-photon_2024}%
  \BibitemOpen
  \bibfield  {author} {\bibinfo {author} {\bibfnamefont {C.}~\bibnamefont
  {Groiseau}}, \bibinfo {author} {\bibfnamefont {A.~I.}\ \bibnamefont
  {Fernández-Domínguez}}, \bibinfo {author} {\bibfnamefont {D.}~\bibnamefont
  {Martín-Cano}}, \ and\ \bibinfo {author} {\bibfnamefont {C.~S.}\
  \bibnamefont {Muñoz}},\ }\href {\doibase 10.1103/PRXQuantum.5.010312}
  {\bibfield  {journal} {\bibinfo  {journal} {PRX Quantum}\ }\textbf {\bibinfo
  {volume} {5}},\ \bibinfo {pages} {010312} (\bibinfo {year}
  {2024})}\BibitemShut {NoStop}%
\bibitem [{\citenamefont {Kutas}\ \emph {et~al.}(2024)\citenamefont {Kutas},
  \citenamefont {Riexinger}, \citenamefont {Klier}, \citenamefont {Molter},\
  and\ \citenamefont {von Freymann}}]{kutas_terahertz_2024}%
  \BibitemOpen
  \bibfield  {author} {\bibinfo {author} {\bibfnamefont {M.}~\bibnamefont
  {Kutas}}, \bibinfo {author} {\bibfnamefont {F.}~\bibnamefont {Riexinger}},
  \bibinfo {author} {\bibfnamefont {J.}~\bibnamefont {Klier}}, \bibinfo
  {author} {\bibfnamefont {D.}~\bibnamefont {Molter}}, \ and\ \bibinfo {author}
  {\bibfnamefont {G.}~\bibnamefont {von Freymann}},\ }\href {\doibase
  10.48550/arXiv.2408.02531} {\enquote {\bibinfo {title} {Terahertz {Quantum}
  {Imaging}},}\ } (\bibinfo {year} {2024}),\ \bibinfo {note} {arXiv:2408.02531
  [physics, physics:quant-ph]}\BibitemShut {NoStop}%
\bibitem [{\citenamefont {Kimble}(2008)}]{kimble_quantum_2008}%
  \BibitemOpen
  \bibfield  {author} {\bibinfo {author} {\bibfnamefont {H.~J.}\ \bibnamefont
  {Kimble}},\ }\href {\doibase 10.1038/nature07127} {\bibfield  {journal}
  {\bibinfo  {journal} {Nature}\ }\textbf {\bibinfo {volume} {453}},\ \bibinfo
  {pages} {1023} (\bibinfo {year} {2008})}\BibitemShut {NoStop}%
\bibitem [{\citenamefont {Planken}\ \emph {et~al.}(2001)\citenamefont
  {Planken}, \citenamefont {Nienhuys}, \citenamefont {Bakker},\ and\
  \citenamefont {Wenckebach}}]{planken2001measurement}%
  \BibitemOpen
  \bibfield  {author} {\bibinfo {author} {\bibfnamefont {P.~C.}\ \bibnamefont
  {Planken}}, \bibinfo {author} {\bibfnamefont {H.-K.}\ \bibnamefont
  {Nienhuys}}, \bibinfo {author} {\bibfnamefont {H.~J.}\ \bibnamefont
  {Bakker}}, \ and\ \bibinfo {author} {\bibfnamefont {T.}~\bibnamefont
  {Wenckebach}},\ }\href@noop {} {\bibfield  {journal} {\bibinfo  {journal}
  {Journal of the Optical Society of America B}\ }\textbf {\bibinfo {volume}
  {18}},\ \bibinfo {pages} {313} (\bibinfo {year} {2001})}\BibitemShut
  {NoStop}%
\bibitem [{\citenamefont {Fujiwara}\ \emph {et~al.}(2007)\citenamefont
  {Fujiwara}, \citenamefont {Maruyama}, \citenamefont {Sugisaki}, \citenamefont
  {Takahashi}, \citenamefont {Aoshima}, \citenamefont {Cogdell},\ and\
  \citenamefont {Hashimoto}}]{fujiwara_determination_2007}%
  \BibitemOpen
  \bibfield  {author} {\bibinfo {author} {\bibfnamefont {M.}~\bibnamefont
  {Fujiwara}}, \bibinfo {author} {\bibfnamefont {M.}~\bibnamefont {Maruyama}},
  \bibinfo {author} {\bibfnamefont {M.}~\bibnamefont {Sugisaki}}, \bibinfo
  {author} {\bibfnamefont {H.}~\bibnamefont {Takahashi}}, \bibinfo {author}
  {\bibfnamefont {S.-i.}\ \bibnamefont {Aoshima}}, \bibinfo {author}
  {\bibfnamefont {R.~J.}\ \bibnamefont {Cogdell}}, \ and\ \bibinfo {author}
  {\bibfnamefont {H.}~\bibnamefont {Hashimoto}},\ }\href {\doibase
  10.1143/JJAP.46.1528} {\bibfield  {journal} {\bibinfo  {journal} {Jpn. J.
  Appl. Phys.}\ }\textbf {\bibinfo {volume} {46}},\ \bibinfo {pages} {1528}
  (\bibinfo {year} {2007})}\BibitemShut {NoStop}%
\bibitem [{\citenamefont {Zhou}\ \emph {et~al.}(2020)\citenamefont {Zhou},
  \citenamefont {Alam}, \citenamefont {Karimi}, \citenamefont {Upham},
  \citenamefont {Reshef}, \citenamefont {Liu}, \citenamefont {Willner},\ and\
  \citenamefont {Boyd}}]{zhou_broadband_2020}%
  \BibitemOpen
  \bibfield  {author} {\bibinfo {author} {\bibfnamefont {Y.}~\bibnamefont
  {Zhou}}, \bibinfo {author} {\bibfnamefont {M.~Z.}\ \bibnamefont {Alam}},
  \bibinfo {author} {\bibfnamefont {M.}~\bibnamefont {Karimi}}, \bibinfo
  {author} {\bibfnamefont {J.}~\bibnamefont {Upham}}, \bibinfo {author}
  {\bibfnamefont {O.}~\bibnamefont {Reshef}}, \bibinfo {author} {\bibfnamefont
  {C.}~\bibnamefont {Liu}}, \bibinfo {author} {\bibfnamefont {A.~E.}\
  \bibnamefont {Willner}}, \ and\ \bibinfo {author} {\bibfnamefont {R.~W.}\
  \bibnamefont {Boyd}},\ }\href {\doibase 10.1038/s41467-020-15682-2}
  {\bibfield  {journal} {\bibinfo  {journal} {Nat. Commun.}\ }\textbf {\bibinfo
  {volume} {11}},\ \bibinfo {pages} {2180} (\bibinfo {year}
  {2020})}\BibitemShut {NoStop}%
\bibitem [{\citenamefont {Giorgianni}\ \emph {et~al.}(2019)\citenamefont
  {Giorgianni}, \citenamefont {Puc}, \citenamefont {Jazbinsek}, \citenamefont
  {Cea}, \citenamefont {Koo}, \citenamefont {Han}, \citenamefont {Kwon},\ and\
  \citenamefont {Vicario}}]{giorgianni_supercontinuum_2019}%
  \BibitemOpen
  \bibfield  {author} {\bibinfo {author} {\bibfnamefont {F.}~\bibnamefont
  {Giorgianni}}, \bibinfo {author} {\bibfnamefont {U.}~\bibnamefont {Puc}},
  \bibinfo {author} {\bibfnamefont {M.}~\bibnamefont {Jazbinsek}}, \bibinfo
  {author} {\bibfnamefont {T.}~\bibnamefont {Cea}}, \bibinfo {author}
  {\bibfnamefont {M.-J.}\ \bibnamefont {Koo}}, \bibinfo {author} {\bibfnamefont
  {J.-H.}\ \bibnamefont {Han}}, \bibinfo {author} {\bibfnamefont {O.-P.}\
  \bibnamefont {Kwon}}, \ and\ \bibinfo {author} {\bibfnamefont
  {C.}~\bibnamefont {Vicario}},\ }\href {\doibase 10.1364/OL.44.004881}
  {\bibfield  {journal} {\bibinfo  {journal} {Opt. Lett.}\ }\textbf {\bibinfo
  {volume} {44}},\ \bibinfo {pages} {4881} (\bibinfo {year}
  {2019})}\BibitemShut {NoStop}%
\bibitem [{\citenamefont {Shields}\ \emph {et~al.}(2022)\citenamefont
  {Shields}, \citenamefont {Dada}, \citenamefont {Hirsch}, \citenamefont
  {Yoon}, \citenamefont {Weaver}, \citenamefont {Faccio}, \citenamefont
  {Caspani}, \citenamefont {Peccianti},\ and\ \citenamefont
  {Clerici}}]{shields_electro-optical_2022}%
  \BibitemOpen
  \bibfield  {author} {\bibinfo {author} {\bibfnamefont {T.}~\bibnamefont
  {Shields}}, \bibinfo {author} {\bibfnamefont {A.~C.}\ \bibnamefont {Dada}},
  \bibinfo {author} {\bibfnamefont {L.}~\bibnamefont {Hirsch}}, \bibinfo
  {author} {\bibfnamefont {S.}~\bibnamefont {Yoon}}, \bibinfo {author}
  {\bibfnamefont {J.~M.~R.}\ \bibnamefont {Weaver}}, \bibinfo {author}
  {\bibfnamefont {D.}~\bibnamefont {Faccio}}, \bibinfo {author} {\bibfnamefont
  {L.}~\bibnamefont {Caspani}}, \bibinfo {author} {\bibfnamefont
  {M.}~\bibnamefont {Peccianti}}, \ and\ \bibinfo {author} {\bibfnamefont
  {M.}~\bibnamefont {Clerici}},\ }\href {\doibase 10.3390/s22239432} {\bibfield
   {journal} {\bibinfo  {journal} {Sensors}\ }\textbf {\bibinfo {volume}
  {22}},\ \bibinfo {pages} {9432} (\bibinfo {year} {2022})}\BibitemShut
  {NoStop}%
\bibitem [{\citenamefont {Gingras}\ and\ \citenamefont
  {Cooke}(2017)}]{Gingras:17}%
  \BibitemOpen
  \bibfield  {author} {\bibinfo {author} {\bibfnamefont {L.}~\bibnamefont
  {Gingras}}\ and\ \bibinfo {author} {\bibfnamefont {D.~G.}\ \bibnamefont
  {Cooke}},\ }\href {\doibase 10.1364/OPTICA.4.001416} {\bibfield  {journal}
  {\bibinfo  {journal} {Optica}\ }\textbf {\bibinfo {volume} {4}},\ \bibinfo
  {pages} {1416} (\bibinfo {year} {2017})}\BibitemShut {NoStop}%
\bibitem [{\citenamefont {Gingras}\ \emph {et~al.}(2018)\citenamefont
  {Gingras}, \citenamefont {Cui}, \citenamefont {Schiff-Kearn}, \citenamefont
  {M\'{e}nard},\ and\ \citenamefont {Cooke}}]{Gingras:18}%
  \BibitemOpen
  \bibfield  {author} {\bibinfo {author} {\bibfnamefont {L.}~\bibnamefont
  {Gingras}}, \bibinfo {author} {\bibfnamefont {W.}~\bibnamefont {Cui}},
  \bibinfo {author} {\bibfnamefont {A.~W.}\ \bibnamefont {Schiff-Kearn}},
  \bibinfo {author} {\bibfnamefont {J.-M.}\ \bibnamefont {M\'{e}nard}}, \ and\
  \bibinfo {author} {\bibfnamefont {D.~G.}\ \bibnamefont {Cooke}},\ }\href
  {\doibase 10.1364/OE.26.013876} {\bibfield  {journal} {\bibinfo  {journal}
  {Opt. Express}\ }\textbf {\bibinfo {volume} {26}},\ \bibinfo {pages} {13876}
  (\bibinfo {year} {2018})}\BibitemShut {NoStop}%
\bibitem [{\citenamefont {Cui}\ \emph {et~al.}(2023)\citenamefont {Cui},
  \citenamefont {Yalavarthi}, \citenamefont {Radhan}, \citenamefont
  {Bashirpour}, \citenamefont {Gamouras},\ and\ \citenamefont
  {Ménard}}]{cui_high-field_2023}%
  \BibitemOpen
  \bibfield  {author} {\bibinfo {author} {\bibfnamefont {W.}~\bibnamefont
  {Cui}}, \bibinfo {author} {\bibfnamefont {E.~K.}\ \bibnamefont {Yalavarthi}},
  \bibinfo {author} {\bibfnamefont {A.~V.}\ \bibnamefont {Radhan}}, \bibinfo
  {author} {\bibfnamefont {M.}~\bibnamefont {Bashirpour}}, \bibinfo {author}
  {\bibfnamefont {A.}~\bibnamefont {Gamouras}}, \ and\ \bibinfo {author}
  {\bibfnamefont {J.-M.}\ \bibnamefont {Ménard}},\ }\href {\doibase
  10.1364/OE.496855} {\bibfield  {journal} {\bibinfo  {journal} {Opt. Express}\
  }\textbf {\bibinfo {volume} {31}},\ \bibinfo {pages} {32468} (\bibinfo {year}
  {2023})}\BibitemShut {NoStop}%
\end{thebibliography}

%

\end{document}